\RequirePackage{fix-cm}
\documentclass[smallextended]{svjour3}
\usepackage{graphicx}
\usepackage{color}
\usepackage{mathptmx}
\usepackage{latexsym}
\journalname{Foundations of Physics} 

\title{Particles, Cutoffs and Inequivalent Representations}
\subtitle{Fraser and Wallace on Quantum Field Theory}

\author{Matthias Egg \and Vincent Lam \and Andrea Oldofredi}

\institute{Matthias Egg \at
              University of Bern, Institute of Philosophy, Laenggassstrasse 49a, Postfach, 3000 Bern 9, Switzerland\\
               \email{matthias.egg@philo.unibe.ch}
             \and
               Vincent Lam \at
               University of Geneva, Department of Philosophy, Rue de Candolle 2, 1211 Gen\`eve 4, Switzerland\\
               \email{vincent.lam@unige.ch}\\
               \& \\
               The University of Queensland, School of Historical and Philosophical Inquiry, St Lucia QLD 4072, Australia\\
               \email{v.lam@uq.edu.au}
             \and
               Andrea Oldofredi \at
               University of Lausanne, Department of Philosophy, 1015 Dorigny, Lausanne, Switzerland\\
               \email{andrea.oldofredi@unil.ch}  
               }
            
\date{Received: date/ Accepted: date}
 
\begin{document}

\maketitle 

\begin{abstract}
We critically review the recent debate between Doreen Fraser and David Wallace on the interpretation of quantum field theory, with the aim of identifying where the core of the disagreement lies. We show that, despite appearances, their conflict does not concern the existence of particles or the occurrence of unitarily inequivalent representations. Instead, the dispute ultimately turns on the very definition of what a quantum field theory is. We further illustrate the fundamental differences between the two approaches by comparing them both to the Bohmian program in quantum field theory.
\keywords{Algebraic quantum field theory \and Particle physics \and Renormalization \and Unitarily inequivalent representations}
\end{abstract}

\section{Introduction}
\label{section 1}

The interpretation of relativistic quantum field theory (QFT) is a difficult task, since it involves specific field-theoretic and relativistic issues on top of the usual quantum puzzles found in non-relativistic quantum mechanics (QM), such as the measurement problem and non-locality. (Merely moving from QM to QFT obviously does not solve the latter puzzles, so any realist interpretation of QFT will have to provide an account of them.) Similarly to the QM case (to some extent), different stances on these QFT issues lead to different variants of the theory. Doreen Fraser and David Wallace have recently debated over the best approach to QFT for foundational and interpretative work, the former arguing in favor of the mathematically rigorous and physically ambitious approach of algebraic quantum field theory (AQFT), the latter defending the more pragmatic approach of `conventional' quantum field theory (CQFT), which is described in most QFT textbooks and used by most working physicists in the domain. The debate between Fraser and Wallace has the great merit of highlighting and focusing on some of the foundational and interpretative issues that appear within the QFT framework, but which have been far less investigated than the usual ones already present in QM. The upshot of their debate for the interpretation of QFT is, however, far from clear and even somewhat confusing. 

In this paper, we critically review this debate, mainly as expressed in \cite{Fraser2009}, \cite{Fraser2011} and \cite{Wallace2006} \cite{Wallace2011}, hopefully clarifying along the way several important points for the interpretation of QFT. In section \ref{section 2}, we will argue that, despite appearances, the disagreement between Fraser and Wallace is not really concerned with the existence of particles (or quanta). We will then progressively approach what we claim to be the core of their disagreement, discussing first their diverging interpretations of renormalization (section \ref{section 3}) and thereafter their differing views on the permissibility of cutoffs (section \ref{section 4}). This crucial issue is intimately connected to the definition of the very project of QFT, which we will discuss in section \ref{section 5}. In section \ref{section 6}, we will spell out what we take to be the upshot of the debate for the foundation of QFT.

\section{Where the conflict does not lie: Particle ontology}
\label{section 2}

One of the most discussed interpretative issues of QFT concerns the extent to which the theory can be understood in terms of particles (or in terms of the weaker notion of quanta). A superficial look at the debate between Fraser and Wallace could lead to the impression that one of their main points of disagreement precisely is the particle interpretation of QFT. After all, one of Fraser's central claims is that a particle ontology is incompatible with the theoretical principles of (any conceptually rigorous) QFT (see especially her \cite{Fraser2008} paper), while Wallace's defense of CQFT seems to support just such an ontology, by arguing for the conceptual respectability of the standard model of particle physics.

On closer inspection, however, this conflict turns out to be unsubstantial, because Fraser and Wallace are actually speaking about different things. What  Fraser \cite{Fraser2008} criticizes is the idea that QFT could be interpreted in terms of a \emph{fundamental} particle notion, or, more precisely, in terms of quanta, that is, countable entities allowing physical states to be characterized by well-defined Fock space occupation numbers. Indeed, she rightly argues that a Fock space representation is in general and strictly speaking only available for free, non-interacting QFT systems; central to her argument is the fact that free and interacting QFT systems involve unitarily inequivalent representations (this can also be seen as a consequence of Haag's theorem). 

Does Wallace \cite{Wallace2011} have in mind arguing for a fundamental ontology of particles when he speaks of ``taking particle physics seriously''? Clearly not. For one thing, he does not deny the foundational importance of the existence of (`infra-red')  unitarily inequivalent representations (\cite[123-124]{Wallace2011}; we will come back to the role unitarily inequivalent representations play in the debate below). More importantly, he does not criticize Fraser's anti-quanta argument \emph{per se}, but only the conclusion she draws from it as regards the question whether there could be approximate agreement on matters of ontology between CQFT and (a possible future interacting model of) AQFT. Fraser has given a negative answer to that question, insisting that ``a theory according to which quanta exist is not approximately equivalent to a theory according to which quanta do not exist'' \cite[560]{Fraser2009}. Wallace replies:

\begin{quote}
Well, maybe not, but it is approximately equivalent to a theory according to which quanta do approximately exist! And if AQFT (more precisely, if this supposed interacting algebraic quantum field theory) does not admit quanta in at least some approximate sense, then so much the worse for it: the evidence for the electron is reasonably conclusive. \cite[123]{Wallace2011}
\end{quote}

This shows that Wallace is not in the business of defending a fundamental particle interpretation of QFT. He is merely committed to the ``approximate existence'' of particles, which means that particles are not basic entities in their own right, but ``just certain patterns of excitations in the [quantum] fields'' \cite[73]{Wallace2006}.\footnote
{Wallace and Timpson \cite[707]{Wallace2010} also emphasize the non-fundamental character of particles in QFT. In contrast to \cite{Wallace2006,Wallace2011}, they even seem to advocate AQFT as a guide to fundamental ontology (711-712). This reinforces the claim that an AQFT-fueled refusal of particles at the fundamental level is fully compatible with a commitment to particles as non-fundamental entities.} Therefore, even if one takes CQFT to be the best theory of the relevant domain (and accepts that we should be ontologically committed to the entities that are said to exist according to our best scientific theories), the resulting fundamental ontology will not be one of particles.\footnote
{\label{field ontology}
Note that, despite its name, QFT does not straightforwardly support a field ontology either \cite{Baker2009}.} We thus have an argument against the fundamentality of particles which is internal to CQFT. 

Furthermore, there is a second, more general argument, which is external to CQFT. It derives from the fact that CQFT, by its very nature, cannot even pretend to be a fundamental theory: as we will see below, CQFT generally goes together with a short-distance cutoff assumption (it makes the theory far better defined by `freezing out' the high energy or short lenghtscale degrees of freedom that are responsible for divergent behavior), which is naturally understood as indicating the breakdown of the theory at some short lengthscales (renormalization theory then ensures the epistemic innocuousness of this assumption). This is usually expressed by saying that CQFT is an \emph{effective} (as opposed to fundamental) theory. \cite[46-50]{Wallace2006} illustrates the status of CQFT by means of an analogy to classical mechanics (CM) and its relation to non-relativistic QM. When one approaches the classical limit of QM, one is considering a certain regime where CM is approximately isomorphic to QM, in the sense that predictions of CM are reasonably good approximations, or isomorphic to, those of the lower level theory, QM. Wallace now applies the same argumentative structure to CQFT, assuming that it breaks down at some small length scale where another, as-yet-unknown physical theory X is supposed to come into the scene. Then, there should be an isomorphism between the two, in the same way as there is a certain regime where CM is approximately isomorphic to QM, or in other words, there should be a particular length-scale above which the predictions of CQFT are good approximations to those of the theory X. According to Wallace, then, the ontology of CQFT is the kind of pattern or structure that emerges from the deeper theory X. Therefore, even if (contrary to what we saw in the previous paragraph) the basic ontology of CQFT included particles, particles would still not be truly fundamental, due to the inherently non-fundamental character of CQFT.\footnote{This does not mean, however, that no ontological lessons can be drawn from an effective theory like CQFT. We will return to this point in section \ref{section 5}.}

To sum up, Wallace clearly does not advocate the kind of (fundamental) particle ontology that Fraser primarily attacks. However, the following passage shows that she takes issue even with his modest commitment to (non-fundamental) particles:

\begin{quote}
Wallace considers the `particle' concept to be an emergent concept [\ldots]. For this to be a viable response, the cogency of the distinction between fundamental and less fundamental entities must be defended and a case must be made for admitting additional, non-fundamental entities into our ontology. \cite[858]{Fraser2008}
\end{quote}

But this obviously is a general concern about non-fundamental entities; it applies to molecules, organisms, or planets no less than to particles (in Wallace's sense). If \emph{that} were the main point of conflict between Fraser and Wallace, it would be hard to understand why their debate should revolve around quantum field theory as much as it does. Therefore, the existence of particles cannot be what this debate is about; ultimately, both Fraser and Wallace---relying on arguments from AQFT and CQFT respectively---agree about particles being non-fundamental in QFT.

\section{Renormalization: Field-theoretic breakdown or underdetermination?}
\label{section 3}

What, then, is the core of the disagreement between Fraser and Wallace? On a purely methodological level, the central question is easily identified by looking at the abstracts of \cite{Fraser2011} and \cite{Wallace2011}: is it AQFT or CQFT that should be subject to foundational analysis? Now the previous section has shown that Fraser's and Wallace's disagreement on this methodological question need not imply a disagreement on matters of ontology, because AQFT and CQFT simply address different kinds of ontological questions: while the former is concerned with \emph{fundamental} ontology, the latter yields what could be called an \emph{effective} ontology. 

However, there is still a substantial point of ontological disagreement, which has to do with the interpretation of renormalization techniques, and in particular, the applicability of field theoretic concepts at lengthscales beyond the cutoff length. Consider the following argument by Wallace:

\begin{quote}
\begin{enumerate}
\item We have a very well confirmed theory (the Standard Model, understood as a CQFT), one of whose central claims is that field degrees of freedom are frozen out at sufficiently short lengthscales.
\item As good scientific realists, we should tentatively accept that claim as approximately true.
\item Since AQFT denies that claim, its basic structure is wrong.
\item So no wonder that we can't construct empirically adequate theories within that structure! \cite[120]{Wallace2011}
\end{enumerate}
\end{quote} 

In her response, Fraser \cite[133]{Fraser2011} criticizes the move from 1 to 2. This move, she claims, is based on the so-called `no miracles' argument for scientific realism, according to which the empirical success of our well-confirmed scientific theories would be inexplicable unless these theories were approximately true. Against this, Fraser advances the argument from underdetermination, which states that, if mutually incompatible theories make the same empirical predictions, empirical success does not support the claim that any particular one of them is approximately true.

Before discussing the different roles that renormalization plays in these two opposing arguments, let us note that both of them have an internal weakness, which is acknowledged by their respective proponents. On the one hand, Wallace admits that his move from 1 to 2 is a little too quick, because ``CQFT, in itself, doesn't actually \emph{require} the existence of a cutoff; it just tells us that \emph{assuming} a cutoff suffices to make the theory well-defined'' \cite[120]{Wallace2011}. Indeed, one of the key insights of modern renormalization theory (repeatedly stressed by both Wallace and Fraser) is that the empirical content of CQFT is largely insensitive to what precisely happens at small lengthscales. Therefore, ``it is consistent with the CQFT framework that the theory's degrees of freedom after all remain defined on arbitrarily short lengthscales'' (ibid.). On the other hand, Fraser has to acknowledge that there is no actual underdetermination between AQFT and CQFT, because so far, only the latter makes empirical predictions about interacting systems in four spacetime dimensions \cite[132]{Fraser2011}. We are thus at present confronted with merely a \emph{potential} underdetermination, premised on the assumption that one day a realistic AQFT model will be found.

That the interpretation and the implications of the renormalization group methods constitute a central point of disagreement in the Fraser-Wallace debate is highlighted by the opposed role they play in the above no miracles argument by Wallace and in Fraser's main objection to it. According to Wallace, renormalization theory removes all the motivation for developing realistic AQFT models in the domain of CQFT (i.e. well above the short lengthscale cutoff), since the renormalization group methods show that the (hypothetical) AQFT models will have the same empirical predictions as CQFT (in the relevant domain). So, even if CQFT is strictly speaking consistent with a full field-theoretic description at all lenghtscales, this possibility does not really matter, thanks to renormalization. However, according to Fraser, this constitutes a clear case of theory underdetermination by empirical evidence, so that she regards renormalization theory as supporting her objection to Wallace's no miracles argument. 

At this stage, two important remarks are in order. First, let us repeat that, as Wallace \cite[121]{Wallace2011} clearly points out, there is no actual underdetermination, since there is no AQFT model reproducing the empirical predictions of CQFT (e.g. the Standard Model) for the time being. Second, as Fraser \cite[131]{Fraser2011} argues, there might be some other motivations besides novel empirical predictions for looking for an alternative to CQFT (even in the domain well above the short length scale cutoff), such as coherence with certain theoretical (e.g. relativistic) principles or ontological clarity.\footnote{\cite{Kuhlmann2010} also privileges AQFT on ontological grounds.} (For instance, this latter is the principal motivation behind Bohmian mechanics, which makes the same predictions as `conventional' or textbook QM; indeed, Wallace \cite[ftn. 14]{Wallace2011}, who privileges the Everett interpretation, seems to recognize that his point about renormalization might be weakened in the cases of the other main realist approaches to the measurement problem such as Bohmian mechanics or dynamical collapse approaches, e.g. GRW. We will return to this point in section \ref{section 5} below.)

All this tends to show that both Wallace's no miracles argument and Fraser's argument from underdetermination---as well as the respective roles of renormalization theory in these arguments---are inconclusive. This might also show that their divergent interpretation of the renormalization group methods---however central to the Fraser-Wallace debate---does not constitute the real bottom line of the discussion. Indeed, as explicitly recognized by both \cite[127]{Fraser2011} and \cite[121]{Wallace2011}, what is really at stake is the status of cutoffs in QFT, to which we now turn.

\section{The status of cutoffs}
\label{section 4}

An important source of inspiration (and justification) for renormalization methods in CQFT is the successful application of such methods in condensed matter physics, where there is good reason to assume that the field theoretic treatment fails at lengthscales small enough for the atomic structure of matter to become relevant. Wallace claims that ``nothing prevents us telling \emph{exactly the same story} in particle physics, provided only that something freezes out the short-distance degrees of freedom on some lengthscale far below what current experimental physics can probe'' \cite[118]{Wallace2011}. In this line of thought, assuming the short-distance breakdown of QFT, i.e. assuming a short-distance cutoff, is a crucial step in the renormalization process and constitutes therefore a key element of the conceptual grounding and of the explanatory framework of CQFT:

\begin{quote}
This, in essence, is how modern particle physics deals with the renormalization problem [footnote deleted]: it is taken to presage an ultimate failure of quantum field theory at some short lengthscale, and once the bare existence of that failure is appreciated, the whole of renormalization theory becomes unproblematic, and indeed predictively powerful in its own right. \cite[119]{Wallace2011} 
\end{quote}

However, Fraser detects an important disanalogy between condensed matter physics and CQFT: ``in the condensed matter case, independent evidence for the existence of atoms plays a pivotal role. [\ldots] There is no analogue of this evidence in the QFT case; we do not possess evidence of this sort that QFT breaks down at short distance scales'' \cite[118]{Wallace2011}.

Actually, Wallace does cite some evidence for an eventual breakdown of QFT at short distance scales:

\begin{quote}
Once we get down to Planckian lengthscales, the fiction that spacetime is nondynamical and that gravity can be ignored will become unsustainable. Whatever our sub-Planckian physics looks like (string theory? twistor theory? loop quantum gravity? non-commutative geometry? causal set theory? something as-yet-undreamed-of?) there are pretty powerful reasons \emph{not} to expect it to look like quantum field theory on a classical background spacetime. \cite[120-121]{Wallace2011}
\end{quote}

Presumably, Fraser does not accept this as evidence in the same sense as the evidence for the existence of atoms, because it is not \emph{experimental} evidence. Now we are not opposed to the idea that experimental evidence should be given more weight than theoretical evidence,\footnote
{
See \cite{Egg2012jgps}, \cite{Egg2014} for a recent version of scientific realism based on this idea. Section 9.3 of the latter work also contains a deeper investigation of the interplay between experimental and theoretical considerations in the interpretation of QFT.
}
but one must be careful not to exaggerate the privilege granted to experimental evidence, because such an attitude (exemplified by \cite{MacKinnon2008}) would undermine the motivation of giving any relevance to the results of AQFT in the first place. Instead, theoretical considerations of the type cited by Wallace should be taken seriously, especially in contexts where no direct experimental evidence is available, as is the case for the question that concerns us here (the breakdown of QFT at small lengthscales).

Once this is acknowledged, the theoretical evidence in favor of a breakdown of QFT at the Planck scale can be evaluated against the (equally theoretical) evidence in favor of the basic commitments of AQFT, which include the claim that a field theoretic treatment is adequate for all lengthscales. Fraser's general argument for accepting the ontological commitments of AQFT (and disregarding those of CQFT) ``is that QFT is a unification project; the goal of the project is to formulate a theory that incorporates both special relativistic and quantum principles'' \cite[131]{Fraser2011}.\footnote
{
Fraser might disagree with our characterization of this kind of evidence as \emph{theoretical}, as she claims that ``this unification project is also empirical, broadly construed, insofar as there is indirect empirical support for special relativity and its theoretical principles and for non-relativistic quantum theory and its theoretical principles'' \cite[131]{Fraser2011}. However, given that there is indirect empirical support for general relativity as well, Wallace's above-mentioned arguments also count as ``empirical'' in this sense.}
She goes on to claim that AQFT satisfies the minimal criteria for success in this unification project (obviously, since it is one of its explicit aims), but CQFT does not (CQFT with cutoff is strictly speaking not relativistic, because it is not Poincar\'e covariant; see section \ref{section 5} below).

We are thus faced with two conflicting bodies of theoretical evidence concerning the breakdown of QFT at short lengthscales, and both of them are based on considerations of unification: unification of quantum theory and special relativity in Fraser's case, unification of quantum theory and general relativity in Wallace's case. The crucial difference is that Fraser takes the unification of quantum theory and special relativity to be part of the very definition of what QFT is, and therefore to be privileged over other considerations, such as gravitational considerations, that are ``external'' to the QFT project \cite[552]{Fraser2009}. By contrast, Wallace mentions the unification of quantum theory and general relativity precisely to show that the search for a truly fundamental theory has to move beyond the QFT framework.

This actually touches the heart of the disagreement between Fraser and Wallace. Since Fraser primarily considers QFT as the project of combining the principles of quantum theory with those of special relativity, she cannot accept cutoffs since they destroy Poincar\'e covariance and thereby contradict the very idea of this project. (She might be willing to accept them in the last resort, after it has been shown that the QFT project as she understands it fails). On the other hand, according to Wallace, such in principle understanding of QFT is illegitimate in the face of its failure (for the time being) to account for the QFT phenomenology (i.e. the Standard Model). By contrast, this latter is naturally accounted for within the framework of CQFT understood as an effective theory breaking down at short lenthscales, and possibly approximating some deeper, more fundamental theory (e.g. one that would take quantum gravitational effects into account). In this context, assuming a short-distance cutoff implements this effective understanding, which can well be motivated by `external' considerations. Renormalization theory ensures that the details of this implementation are irrelevant for the predictive and explanatory power of CQFT (\cite{Wallace2006} alludes to some broad structuralist understanding of theories in order to justify the explanatory power of an approximate theory like CQFT). 

Therefore, we see that Fraser and Wallace ultimately disagree about what QFT is, and this disagreement manifests itself in their opposite understanding of the short-distance cutoff. Before continuing our discussion of their disagreement about the very nature of QFT, we end this section by highlighting an important point of agreement as regards the cutoffs. Somewhat paradoxically at first sight, Fraser and Wallace rejoin on the status of the long-distance (or `infra-red') cutoff. Indeed, if Fraser holds both short- and long-distance cutoffs as unsatisfactory, Wallace seems to think that the (external, mainly cosmological) motivations for the latter are far weaker than for the former (observational data, i.e. experimental evidence, actually seems to speak in favor of an open, infinite universe, that is, against any genuine long-distance cutoff). As a consequence, Wallace 
seriously considers the possibility of the existence of infinitely many QFT degrees of freedom and therefore of inequivalent representations. (Since  the Stone-von Neumann theorem does not apply to infinite dimensional cases, the quantization of a theory with infinitely many degrees of freedom leads to many unitarily inequivalent representations of the relevant algebra encoding the appropriate commutation relations.) It is interesting to note that the existence of inequivalent representations precisely constitutes one of the original motivations for AQFT, which focuses on the algebraic rather than representational structures of the theory. In short, the foundational importance of unitarily inequivalent representations is part of the common ground shared by both Fraser and Wallace. We will come back to this point in section \ref{section 6}.           

\section{What exactly is QFT?}
\label{section 5}

As we have seen in the previous section, Fraser intends QFT as the theory that best unifies the principles of quantum mechanics and special relativity (SR). From this naturally follows the definition of what QFT should be: in order to be an acceptable QFT, a theory must incorporate the principles of both quantum mechanics and special relativity, in short QFT = QM + SR. This definition finds its roots in the history of quantum field theory,\footnote{The reader interested in historical aspects of the theory should refer to \cite{Schweber:1994aa} and \cite{Cao:1997aa}} since already from the Twenties and the Thirties physicists were trying to incorporate within the domain of quantum mechanics the treatment of the electromagnetic field, given the fact that standard QM does not provide any treatment of relativistic particles. Thus, a quantum theory able to describe these particles was needed, and QFT is the first theory which yields this results, providing a successful treatment of photons. In addition to these historical considerations that Fraser considers to justify her definition, it should be noted that it is congenial to her argument, since among the variants of QFT she considers only AQFT satisfies it, being a project directly aimed to find a consistent (explicitly) relativistic QFT. Certainly this fact plays a pivotal role in arguing why one should prefer the axiomatic variant over the rivals.

Showing how the interaction picture is inconsistent as a consequence of Haag's theorem, Fraser \cite[544-553]{Fraser2009} discusses various modalities to solve the inconsistency: on the one hand we have a formal or `principled' response (AQFT), exemplified 
by the Glimm and Jaffe model, and on the other hand, the approaches, to which CQFT belongs, relying on renormalization methods and in particular on the introduction of cutoffs to solve the issue. The introduction of cutoffs implies that the theory under consideration is not covariant under Poincar\'e transformations and in brief that it is not relativistic.\footnote{See \cite[50-52]{Wallace2006} for different ways to address the problem of Poincar\'e non-covariance from the perspective of CQFT.} According to Fraser's definition of QFT, if a theory is not relativistic, it cannot be a good candidate for a QFT.

Therefore, Fraser does not only claim that we should prefer AQFT for its clarity in solving the inconsistency of the interaction picture, but she also gives another argument, namely that AQFT is genuinely relativistic and thereby, unlike its rivals, fulfills the minimal criterion for being a QFT in the first place. This shows how Fraser's definition of a QFT plays a crucial role in her argument.

A radical consequence of this approach is that a great variety of different formulations and approaches to QFT are simply ruled out in principle. Apart from CQFT, the Bohmian approach to QFT constitutes another good example of such an alternative ruled out from the start (for a recent and accessible overview of this approach to QFT, see \cite{Struyve2011}). This latter approach actually highlights important aspects of the discussion, because the Bohmian versions of QFT share crucial features with both CQFT and AQFT. 

Let us discuss these similarities in turn.

On the one hand, the Bohmian QFTs rely completely on the mathematical structures of CQFT, meaning that all the perturbation and scattering theory on which the standard QFT is based is simply retained without modifications in this theoretical framework. This fact implies that, as in the non-relativistic case, the Bohmian theories add to the standard quantum theory a set of (so-called `primitive') variables and provide their dynamical laws of motion, leaving untouched the mathematical machinery of both standard QM and QFT. The consequence for these Bohmian quantum field theories is that they are not genuinely relativistic and therefore fall outside Fraser's definition of QFT. Furthermore, the fact that these theories rely on the very same regularization techniques as CQFT makes them in some sense explicitly \emph{effective}, rather than \emph{fundamental} theories \cite{Lam2015}. This raises the issue of the ontological impact of (explicitly) effective theories.

According to the Bohmian approach, one should be able to provide a clear (even if non-fundamental) ontological picture for effective theories too. More particularly, such a picture should be able to provide an explicit account of the empirical data that constitute the basis for empirical confirmation, by introducing primitive variables referring to positions in 3-dimensional physical space, the temporal development of which is then described by the theory. (This is the spirit of John S. Bell's quest for `local beables', which is echoed by the so-called `primitive ontology' approach in the current debate on the foundations of quantum theory; see \cite{Allori2015} for an up-to-date review.) 

On the other hand, Bohmian QFTs share with AQFT a striving for the kind of conceptual and ontological clarity which Fraser and others find lacking in CQFT (see section \ref{section 3} above). By a selection of either a particle or a field ontology the Bohmian QFTs aim to construct a theory without ambiguous notions appearing in the axioms and able to solve the measurement problem, which carries over intact in the passage from non-relativistic QM to QFT (for a recent discussion of the measurement problem in QFT see \cite{Barrett:2014aa}), following the strategy applied in the non-relativistic case. These theories are by construction empirically equivalent to any \emph{regularized} QFT, so they are able to provide a qualitative description of the phenomenology of the standard model of particle physics in terms of the motion of local beables, and thus, they are able to explain in a clear manner the phenomena of particle creation and annihilation. Exactly as in the case of AQFT, these Bohmian quantum field theories try to propose a clear solution to the conceptual issues affecting CQFT. They do so by starting from the definition of the local beables of the theory and their respective dynamical equations.

Clearly the Bohmian and the algebraic approaches to QFT aim to solve different problems of CQFT:\footnote{These two approaches have clearly a different scope and in some sense Bohmian QFT is less ambitious than AQFT, but far more empirically successful, since it is built to be empirically equivalent to CQFT as explained above.}
on the one hand, one has the usual problems of QM which are simply inherited by QFT such as the measurement problem and the lack of a clear ontology. These problems imply that QFT is not a theory able to provide a clear description of the physical processes taking place at the subatomic level. For instance, between the free asymptotic states at minus and plus infinity, CQFT does not yield a description of the motion and interactions of the quantum objects going on in scattering processes. Bohmian approaches to QFT aim at addressing these problems. On the other hand, AQFT starts from a rigorous unification of QM and SR trying to construct a theory which is genuinely relativistic and able to remove the cutoffs, giving less importance to the ontological shortcomings of CQFT.

However, one should acknowledge that both these two research programs are worthy of consideration, since they might provide valuable solutions to the wide spectrum of the well-known problems affecting CQFT. More importantly, since these two alternative approaches to QFT rely on different assumptions and solve different problems of the same theory, it would be extremely useful for the future developments in QFT to make an attempt in unifying these two frameworks looking for a QFT with a clear ontology in the spirit of the primitive ontology approach and grounded in a rigorous mathematics, avoiding the ambiguities of the standard approach to QFT.\footnote{A possible proposal could be to cast the existing Bohmian QFTs in the framework of Wightman's axioms as it has been suggested by Nino Zangh\'i (personal communication).}

According to Fraser's approach to QFT, however, whatever the effective ontological clarity provided by the Bohmian versions of QFT, they should be dismissed for foundational work in the QFT domain since they are not relativistic. This straightforward consequence suggests a few critical remarks about Fraser's stance on QFT. First, the Bohmian approach to QFT can be understood as an on-going research program, in some respect with an analogous status to the AQFT research program, since both these approaches to QFT aim to improve the conceptual and technical issues present in the standard formulation of the theory, even though they try to solve different problems and move from diverse metaphysical stances. Furthermore, there is current research in order to develop a relativistic version of Bohmian QFT (see \cite{DGZ2013}) and considerable philosophical work which tries to make clear in which sense this theory can be made compatible with special relativity. To this regard, it is important to stress that BQFTs maintain by construction at least two crucial feature of SR: Bohmian particles\footnote{It is interesting to consider that both in 
Bell's first pilot-wave model (\cite{Bell:1986aa}) as well as in the Dirac Sea formulation of BQFT (\cite{Colin:2007aa}) bosons are not part of the ontology, fermions are sufficient in order to explain and describe observed phenomena. Bosons are instead part of the ontology in the Bell-type QFT, where they receive the same particle status 
as the fermions. For a detailed technical and conceptual expositions of these ideas see \cite{Durr:2005aa}.} do not travel faster than the velocity of light and the quantum equilibrium hypothesis ensures that superluminal signaling is completely avoided. 
Then, the cogent problem of the pilot-wave approach is to find a consistent way to implement Lorentz invariance, and many efforts in this direction have been made (for instance \cite{Durretal.}). Secondly, one has to acknowledge that even if these theories do imply a preferred reference frame, it is a well-argued fact that there is no possibility in principle for an observer to determine it experimentally. The predictions of Bohmian QFTs (just like those of CQFT) are therefore operationally Lorentz invariant, so that the objection from SR (at the heart of the motivation for AQFT) is somewhat weakened.

Finally, and although this is still a hotly debated issue, it is fair to say that Bell's theorem and its consequences on locality should at least inspire some caution as regards the ultimate compatibility of the principles of QM with those of SR (see \cite{Seevinck2010} and 
\cite{Goldstein2011} for recent surveys of this debate, and \cite{Maudlin2011} for an accessible introduction to the tension between QM and SR). In other words, it is still a largely open question whether or not it is even possible to find a theory which combines the principles of both QM and SR. Fraser is not committed to a positive answer, and she stresses the need for further work on the issue: 

\begin{quote}
Since QFT= QT [quantum theory] + SR, the project of formulating quantum field theory cannot be considered successful until either a consistent theory that incorporates both relativistic and quantum principles has been obtained or a convincing argument has been made that such a theory is not possible. The big foundational question lying in the background is, of course, whether the principles of quantum theory and special relativity are consistent. I do not presume that this question has a positive answer. However, I do maintain that the project of developing QFT cannot be considered complete until this central foundational question has been answered. \cite[550]{Fraser2009}
\end{quote} 

Almost anyone interested in the foundations of QFT would concur with Fraser here on the importance of investigating to what extent the principles of quantum theory and special relativity are consistent. But this provides no ground for dismissing alternative developing research programs that do not start with Fraser's strict definition of QFT; they may well be of significant interest from both a physical and philosophical perspective. Thus, as physics stands nowadays (in particular given Bell's theorem), it seems wise not to rule out these different strategies to find a better and more rigorous formulation of QFT in both technical and conceptual sense.          

\section{Upshot of the debate for the foundations of QFT}
\label{section 6}

While the main purpose of our paper has been to identify the precise locus of  disagreement between Fraser and Wallace, we 
have %
also discovered important areas of common ground between them. The first one has to do with the non-fundamental status of particles in QFT (discussed in section \ref{section 2}), the second one with the rejection of a long-distance cutoff and the resulting appearance of unitarily inequivalent representations even in renormalized QFT (discussed at the end of section \ref{section 4}). These two areas are actually connected. To see how, consider the following methodological consequence that Wallace draws from the existence of unitarily inequivalent representations:

\begin{quote}
This is one place where my earlier disclaimer is relevant: this kind of \emph{long}-distance (or `infra-red') divergence arises because the free-field vacuum and the interacting-field vacuum [\ldots] differ on arbitrarily long lenghtscales and so are unitarily inequivalent. This really is a case where algebraic methods are required to understand what is going on. \cite[121]{Wallace2011}
\end{quote}

Wallace's earlier disclaimer concerns the distinction between AQFT and QFT studied by algebraic methods. His point here is clear enough: regarding QFT as an effective theory---hence best characterized by CQFT---does not prevent one from using algebraic methods (and, for that matter, to recognize that the theory possibly contains infinitely many degrees of freedom and has many inequivalent representations). But Wallace actually goes further than that when he claims that his purpose has not been ``to suggest that philosophy of quantum field theory done in the AQFT framework has nothing to teach us. On the contrary, much---perhaps most---of that work probably transfers across just fine to CQFT'' \cite[124]{Wallace2011}. In particular, there are many precise results obtained within the AQFT framework highlighting the difficulties with the notion of particles or quanta in the context of QFT, and Wallace clearly thinks they transfer to CQFT (see his \cite[section 5]{Wallace2006}). So, partly based on these results, and as we have stressed in section \ref{section 2}, neither Wallace nor Fraser (a fortiori) think the notion of particles or quanta is fundamental in any sense. This is an important moral of their debate. 

As a further note, we would like to stress, however, that the impact of this moral actually depends on the ontological meaning one gives to the quantum formalism (algebraic or otherwise) within which these results are phrased---that is, it depends on the solution to the quantum measurement problem one favors. (As already mentioned, Wallace---but not Fraser---is clear about that: he favors the Everett interpretation, within which he is happy to accept that particles or quanta are not fundamental.) For instance, the Bohmian approach to QFT mentioned in section \ref{section 5} does not reject the validity of the above-mentioned AQFT results about particles (they are mostly mathematical results), but it offers an alternative understanding of these results (in particular, in the Bohmian context, quantum vectors and operators do not straightforwardly refer to something in the world, see \cite{Daumer1997}). Of course, to the extent that the Bohmian versions of QFT rely on a Fock space representation, the issue of unitarily inequivalent representations still has to be addressed (see \cite{Lam2015}; this seems to be also valid for the dynamical collapse approach to QFT). Both the measurement problem and the existence of unitarily inequivalent representations need to be taken seriously.     

In our search for the core of the disagreement between Fraser and Wallace, we found that they ultimately disagree about what QFT is. Is the bottom line of this debate therefore merely a verbal point on where (not) to attach the label `QFT'? Certainly not. As we sought to show in sections \ref{section 3} and \ref{section 4}, there are substantial issues about the interpretation of renormalization methods, in particular, the short-distance cutoff implemented in CQFT. Regardless of whether one thinks that this excludes CQFT from the class of genuine QFTs, one should acknowledge the non-fundamental character of CQFT brought out by these methods. Connecting this to our discussion in section \ref{section 5} leads us to a final important lesson from the debate between Fraser and Wallace: at least at the present stage of development, not only fundamental but also effective theories should be allowed to inform our ontology.

\begin{acknowledgement}
VL is grateful to the Swiss National Science Foundation for financial support (project n\textsuperscript{o} 169313).
\end{acknowledgement}

\bibliographystyle{spphys}
\bibliography{QFT}

\end{document}